\documentclass[12pt]{article}%
\usepackage{citesort}
\usepackage{amsmath}
\usepackage{graphicx}%
\usepackage{amsfonts}%
\usepackage{amssymb}
\begin{document}

\title{Measurement of asymmetric component in\\proton-proton collisions }
\author{Adam Bzdak\thanks{e-mail: Adam.Bzdak@ifj.edu.pl}\\Institute of Nuclear Physics, Polish Academy of Sciences\\Radzikowskiego 152, 31-342 Krak\'{o}w, Poland}
\maketitle

\begin{abstract}
It is argued that a standard measurement of multiplicities in proton-proton
collisions is sufficient to construct a single nucleon fragmentation function.
A proposed method is based on measurement of mean values of produced particles
$\left\langle n\right\rangle $ and pairs of particles $\left\langle
n(n-1)\right\rangle $ in symmetric and asymmetric bins.

\vskip 0.6cm


\end{abstract}

\section{Introduction}

It is commonly believed that the spectra of particles produced in inelastic
nucleon-nucleon collisions originate not only from a projectile and a target
fragmentations but also from a central production component \cite{FK}. On the
other hand there are many phenomenological and experimental evidences
supporting the idea of a two-component picture (independent target and
projectile contributions) of soft particle production in hadronic collisions.
One of these is the success of the wounded nucleon model \cite{WNM} in
description of pseudorapidity spectra of charged particles in $dAu$ collisions
measured by the PHOBOS collaboration at $\sqrt{s}=200$ GeV \cite{PHO-dAu}.
Indeed, in this calculation Bia\l as and Czy\.{z} \cite{ff-bc} explicitly
assumed that all particles are produced independently from the left- and
right-moving wounded nucleons. As a result the wounded nucleon fragmentation
function was extracted. Similar analysis was performed in the wounded
quark-diquark model \cite{ff-bb} which resulted in simultaneous description of
the PHOBOS $pp,$ $dAu,$ $CuCu$ and $AuAu$ collisions data at $\sqrt{s}=200$
GeV in almost full pseudorapidity range.

The two-component picture of pion (and baryon) production in hadronic
collisions was also extensively studied at SPS energies \cite{ff-Ryb}. It was
proved to be consistent with many experimental findings which include (i) the
absence of long-range two-particle correlations in $pp$ collisions at
$|x_{F}|>0.2$ (ii) the presence of forward-backward multiplicity correlations
in $pp$ collisions at $|x_{F}|<0.2$ and (iii) the $x_{F}$ dependence of the
$\pi^{+}/\pi^{-}$ ratio in averaged $\pi^{+}p$ and $\pi^{-}p$ collisions. The
wounded nucleon fragmentation function deduced from the above was found to be
in a good qualitative agreement with the one extracted from the analysis of
$dAu$ collisions data in the wounded nucleon (quark-diquark) model
\cite{ff-bc,ff-bb}. Indeed, both functions are peaked in the forward direction
and substantially feed into the opposite hemispheres. This fact also speaks
for the validity of the two-component picture of soft particle production.

Unfortunately, the above-mentioned procedures rely on rarely available precise
experimental data either in nucleon-nucleus (deuter-nucleus) collisions or in
$\pi^{+}p$, $\pi^{-}p$ and $pp$ collisions. Moreover, in the former case we
also rely on the specific model of particle production in such reactions.

In the present paper we show that the particle density from one wounded
nucleon can be extracted solely from appropriate multiplicity measurements in
$pp$ collisions. Our method is based on measurement of average numbers of
produced particles $\left\langle n\right\rangle $ and pairs of particles
$\left\langle n(n-1)\right\rangle $ in different symmetric and asymmetric bins.

In the next section we describe our method in detail. In Section 3 some
comments are included. All calculations are presented in the Appendix.

\section{Measurement}

The measurement of a single wounded nucleon fragmentation function can be
performed as follows.

\begin{figure}[h]
\begin{center}
\includegraphics[scale=1.1]{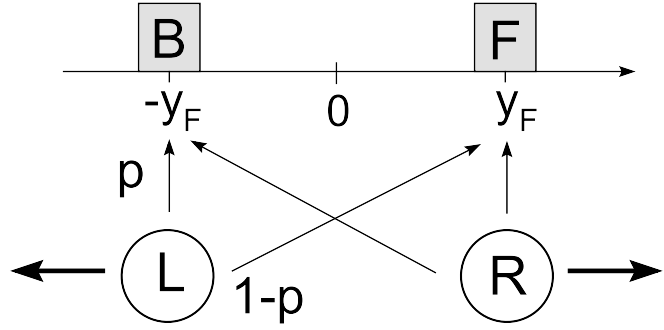}
\end{center}
\caption{In the first step we measure $\left\langle n\right\rangle _{B+F}$ and
$\left\langle n(n-1)\right\rangle _{B+F}$ in the combined interval $B+F$. The
arrows indicate that each wounded nucleon may populate particles into both
intervals.}%
\label{fig_1}%
\end{figure}

In the first step we measure in $pp$ collisions the average numbers of
produced particles $\left\langle n\right\rangle _{B+F}$ and pairs of particles
$\left\langle n(n-1)\right\rangle _{B+F}$ in the combined interval $B+F$,
where $B$ and $F$ are two symmetric (around $y=0$ in the c.m. frame)
rapidity\footnote{Our discussion is valid for any longitudinal variable, not
necessarily rapidity.} intervals. The schematic view of this process is shown
in Fig. \ref{fig_1}. The arrows indicate that the left- and right-moving
wounded nucleons may populate particles into both intervals. $p$ is the
probability that a particle originating from the right(left)-moving wounded
nucleon goes to $F(B)$ interval, under the condition that this particle was
found either in $B$ or $F$.\footnote{Obviously $p$ depends on both position
and width of $F$ interval ($B$ is symmetric around $y=0$).} Consequently, the
probability that a particle originating from the left(right)-moving wounded
nucleon goes to $F(B)$ interval equals $1-p$. 

\begin{figure}[h]
\begin{center}
\includegraphics[scale=1.1]{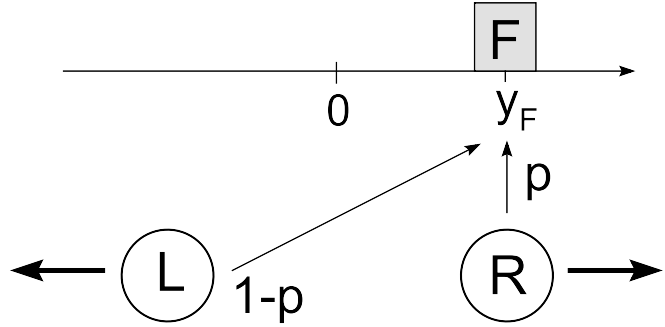}
\end{center}
\caption{In the second step we measure $\left\langle n(n-1)\right\rangle _{F}$
in $F$ interval. Both nucleons may populate to this interval with an
appropriate probabilities $p$ and $1-p$.}%
\label{fig_2}%
\end{figure}

In the second step we repeat the previous measurement but now only in $F$
interval, see Fig. \ref{fig_2}. We measure $\left\langle n(n-1)\right\rangle
_{F}$ and $\left\langle n\right\rangle _{F}$, where of course $\left\langle
n\right\rangle _{F}=\left\langle n\right\rangle _{B+F}/2$. 

Assuming that both nucleons fragment independently we proved, see the
Appendix, that the probability $p$ can be expressed by the previously measured
quantities
\begin{equation}
p=\frac{1}{2}+\frac{1}{2}\sqrt{\frac{4\left\langle n(n-1)\right\rangle
_{F}-\left\langle n(n-1)\right\rangle _{B+F}}{\left\langle n(n-1)\right\rangle
_{B+F}-\left\langle n\right\rangle _{B+F}^{2}}}. \label{p1}%
\end{equation}

Let $\rho_{R}(y)$ be the right-moving wounded nucleon fragmentation
function\footnote{In the c.m. frame a contribution from the left-moving
nucleon $\rho_{L}(y)=\rho_{R}(-y)$.}. It is obvious that the probability $p$
can be expressed as%
\begin{equation}
p=\frac{\int_{F}\rho_{R}(y)dy}{\int_{B+F}\rho_{R}(y)dy}=\frac{\int_{F}\rho
_{R}(y)dy}{\left\langle n\right\rangle _{F}}, \label{p2}%
\end{equation}
where the numerator represents the number of particles in $F$ interval
originating from the right-moving wounded nucleon.

Combining (\ref{p1}) and (\ref{p2}) we obtain the relation between unknown
density of produced particles from a single wounded nucleon (integrated over
$F$) and previously measured $\left\langle n\right\rangle _{B+F}=2\left\langle
n\right\rangle _{F}$, $\left\langle n(n-1)\right\rangle _{B+F}$ and
$\left\langle n(n-1)\right\rangle _{F}$.

Finally, let us notice that if $F$ is sufficiently narrow around $y_{F}$ we
obtain%
\begin{equation}
p=\frac{\int_{F}\rho_{R}(y)dy}{\int_{F}N(y)dy}\approx\frac{\rho_{R}(y_{F}%
)}{N(y_{F})}, \label{p3}%
\end{equation}
which directly relates the fragmentation function $\rho_{R}(y)$ with the
rapidity multiplicity distribution measured in $pp$ collisions $N(y)\equiv
dN(y)/dy$.

\section{Comments}

Following comments are in order.

(i) It is interesting to note that the result (\ref{p1}) does not depend on
the number of active sources of particles (constituents) inside the proton.
Our method rely only on the assumption that both nucleons fragment independently.

(ii) Suppose that the multiplicity distributions in $B+F$ and $F$ intervals
measured in $pp$ collisions can be approximated by the negative binomial
distribution \cite{nbd} with appropriate $k_{B+F}$ and $k_{F}$, where $1/k$
measures the deviation from Poisson distribution. In this case Eq. (\ref{p1})
has a particularly simple form
\begin{equation}
p=\frac{1}{2}+\frac{1}{2}\sqrt{\frac{k_{B+F}}{k_{F}}-1}. \label{p4}%
\end{equation}

(iii) It is enough to perform measurement for different bins up to
$y=y_{F}^{\ast}>0$ in which $p=1$. Here $F$ is populated only by the
right-moving nucleon and $B$ only be the left-moving one, thus $\rho_{R}(y)=$
$N(y)$ for $y\geq y_{F}^{\ast}$. Direct application of \ Eq. (\ref{p1}) will
be disturbed in the far fragmentation region ($y\approx y_{\text{beam}}$),
where energy conservation effects will play an important role.

(iv) Once a single nucleon fragmentation function is measured it can be
directly tested in proton-nucleus collisions (or any other asymmetric
nucleus-nucleus collisions).

(v) Finally, we would like to emphasize that our method apply to any
longitudinal variable, not necessary rapidity.

\bigskip

\textbf{Acknowledgements}

We would like to thank A. Bia\l as and A. Rybicki for useful discussions and
critical reading of the manuscript. This investigation was supported in part
by the Polish Ministry of Science and Higher Education, grant No. N202 034 32/0918.

\appendix                                

\section{Appendix}

Let $P_{B+F}(n)$ be the multiplicity distribution from both wounded nucleons
in the combined interval $B+F$, where $B$ and $F$ are two symmetric rapidity
bins. It is convenient to construct the generating function%
\begin{equation}
H_{B+F}(z)=\sum\nolimits_{n}P_{B+F}(n)z^{n}.
\end{equation}

Let $P_{B+F}^{L}(n)$ and $H_{B+F}^{L}(z)$ be the multiplicity distribution and
corresponding generating function in $B+F$ from the left-moving $L$ wounded
nucleon [and analogous $P_{B+F}^{R}(n)$ and $H_{B+F}^{R}(z)$ for the
right-mover $R$].

Assuming that both nucleons fragment independently we obtain
\begin{equation}
H_{B+F}(z)=H_{B+F}^{L}(z)H_{B+F}^{R}(z). \label{HBF}%
\end{equation}
For collision of two identical nucleons we obtain
\begin{equation}
H_{B+F}^{L}(z)=H_{B+F}^{R}(z)=\sqrt{H_{B+F}(z)}. \label{HLBF}%
\end{equation}

Let $P_{F}^{L}(n)$ be the multiplicity distribution in $F$ interval from the
left-moving nucleon. It can be easily expressed by $P_{B+F}^{L}$
\begin{equation}
P_{F}^{L}(n)=\sum\nolimits_{n^{\prime}\geq n}P_{B+F}^{L}(n^{\prime
})\frac{n^{\prime}!}{n!\left(  n^{\prime}-n\right)  !}\left(  1-p\right)
^{n}p^{n^{\prime}-n},
\end{equation}
where $p$ is the conditional probability that a particle originating from the
left-moving wounded nucleon goes to $B$ interval rather than to $F$.
Consequently $1-p$ is the probability that a particle originating from the
left-moving wounded nucleon goes to $F$ interval rather than to $B$. The
corresponding generating function
\begin{align}
H_{F}^{L}(z)  &  =\sum\nolimits_{n}P_{F}^{L}(n)z^{n}=\sum\nolimits_{n^{\prime
}}P_{B+F}^{L}(n^{\prime})\left[  p+z-pz\right]  ^{n^{\prime}}\nonumber\\
&  =H_{B+F}^{L}(p+z-pz). \label{HLF}%
\end{align}
Performing similar calculations for the right-moving nucleon we obtain%
\begin{equation}
H_{F}^{R}(z)=H_{B+F}^{R}(1-p+pz). \label{HRF}%
\end{equation}

Let $P_{F}(n)$ be the multiplicity distribution from both wounded nucleons in
$F$. Taking Eqs. (\ref{HLBF}), (\ref{HLF}) and (\ref{HRF}) into account the
corresponding generating function $H_{F}$ can be expressed by $H_{B+F}$%
\begin{align}
H_{F}(z)  &  =H_{F}^{L}(z)H_{F}^{R}(z)\nonumber\\
&  =H_{B+F}^{L}(p+z-pz)H_{B+F}^{R}(1-p+pz)\nonumber\\
&  =\sqrt{H_{B+F}(p+z-pz)H_{B+F}(1-p+pz)}. \label{HF}%
\end{align}

Finally, let us calculate the second derivative of Eq. (\ref{HF}) with respect
to $z$ at $z=1$. Taking the following relations into account%
\begin{align}
\left.  \frac{dH_{B+F}(z)}{dz}\right|  _{z=1} &  =\left\langle n\right\rangle
_{B+F},\quad\left.  \frac{d^{2}H_{B+F}(z)}{dz^{2}}\right|  _{z=1}=\left\langle
n(n-1)\right\rangle _{B+F},\nonumber\\
\left.  \frac{d^{2}H_{F}(z)}{dz^{2}}\right|  _{z=1} &  =\left\langle
n(n-1)\right\rangle _{F},
\end{align}
we obtain
\begin{equation}
4\left\langle n(n-1)\right\rangle _{F}=\left\langle n(n-1)\right\rangle
_{B+F}+[1-4p(1-p)]\left[  \left\langle n(n-1)\right\rangle _{B+F}-\left\langle
n\right\rangle _{B+F}^{2}\right]  ,\label{eq}%
\end{equation}
which allows to extract\footnote{We assume that $p\geq1/2$. It corresponds to
the natural assumption that the right-moving nucleon fragmentation function is
peaked in the right hemisphere.} the probability $p$ given by Eq. (\ref{p1}).

\end{document}